\documentclass{aa}  

\usepackage{graphicx}
\usepackage{txfonts}
\begin{document}

   \title{An insight into Capella ($\alpha$ Aurigae): from the extent of core overshoot to its evolutionary history}

\author{E. Marini\inst{1}, C. Ventura\inst{2}, M. Tailo\inst{3}, P. Ventura\inst{1}, F. Dell'Agli\inst{1}, M. Castellani\inst{1}
          }

   \institute{INAF, Observatory of Rome, Via Frascati 33, 00077 Monte Porzio Catone (RM), Italy \\ email: ester.marini@inaf.it \and 
   Dipartimento di Fisica, Sapienza Università di Roma, Piazzale Aldo Moro 5, I-00185 Roma, Italy \and
   Dipartimento di Fisica e Astronomia Augusto Righi, Università degli Studi di Bologna, Via Gobetti 93/2, 40129 Bologna, Italy
   }

   \date{2023}

% \abstract{}{}{}{}{} 
% 5 {} token are mandatory
 
  \abstract
   {The binary star $\alpha$ Aurigae (otherwise known as Capella) is extremely important
   to understand the core hydrogen and helium burning phases of the stars, as the primary star
   is likely evolving through the core helium burning phase, and the masses of the two components
   are $\sim 2.5~{\rm M}_{\odot}$ and $\sim 2.6~{\rm M}_{\odot}$, which fall into a mass range for which the
   extention of the core overshoot during the main sequence phase is uncertain.}
   {We aim at deriving the extent of the core overshoot experienced during the core burning 
    phases and testing the efficiency of the convective transport of energy in the external 
    envelope, by comparing results from stellar evolution modelling with the results from
    the observations.}
   {We consider evolutionary tracks calculated on purpose for the present work, for the primary
    and secondary star of Capella. We determine the extent of the extra-mixing from the core
    during the main sequence evolution and the age of the system, by requiring that the effective
    temperatures and surface gravities of the model stars reproduce those derived from the
    observations at the same epoch. We further check consistency between the observed and
    predicted surface chemistry of the stars.}
   {Consistency between results from stellar evolution modelling and the observations of Capella is
    found when extra-mixing from the core is assumed, the extent of the extra-mixed zone
    being of the order of $0.25~{\rm H_P}$. The age of the system is estimated to be 710 Myr.
    These results allow to nicely reproduce the observed surface chemistry, particularly the
    recent determination of the $^{12}$C$/^{13}$C ratio based on LBT (Large Binocular Telescope) and VATT (Vatican Advanced Technology Telescope) observations.}
   {}


   \titlerunning{Capella: extent of core overshoot and evolutionary history}
   \authorrunning{E. Marini et al.}
   \maketitle
%
%-------------------------------------------------------------------

\section{Introduction}
Understanding the physics of stars demands a thorough comprehension of how
sources of different mass and metallicity evolve, and how the corresponding 
evolutionary tracks move across the Hertzprung-Russell diagram. On this regards
spectroscopic binaries prove a crucial tool, as they allow 
to infer the masses of the individual components with great precision, something
that eases the comparison between the observational evidence and the theoretical
predictions.

Among the various uncertainties still affecting the reliability of results from
stellar evolution modelling, we stress those connected to the convective instability,
namely: a) the extension of the convective core during the main sequence (MS) evolution of
${\rm M} \geq 1.2~{\rm M}_{\odot}$ stars; b) the extent of the mixing occurring during
the red giant branch (RGB) ascending, which brings the surface regions of the stars in contact
with internal zones, previously touched by nuclear activity; c) the efficiency of the
convective transport of energy, which reflects into the overadiabaticity required
in the outer layers of the envelope, thus the temperature gradient in those regions.

The size of the convective core determines the duration of the core hydrogen burning phase, the luminosity 
of the turn-off (the point in the HR diagram where the evolutionary tracks bend to
the red after the consumption of the central hydrogen), and the luminosity of the ''bump'' 
defined by the evolutionary tracks during the core helium burning phase. 
The discussions on this argument date back to the pioneering investigations aimed at
calibrating the extent of overshooting from the convective core of stars evolving through 
the MS \citep{bressan81, renzini87, chiosi92}. The debate was mainly focused on the 
need of assuming some overshoot from the convective core to reproduce the turn off
morphology and luminosity of open clusters: among others, we recall the divergences regarding
the cluster NGC 1866, with some research groups claiming that no-overshoot models were
consistent with the observations \citep{brocato89, testa99}, while other teams stressed the need
of assuming a significant extra-mixing from the core \citep{chiosi89, barmina02}.
Further applications of core overshoot regarded the mass discrepancy problems of
Cepheid stars \citep{bertelli93}.
Recently, binary systems have 
been extensively used to determine the extent of core overshoot as a function of the stellar 
mass \citep{claret17, claret19, morales22}. Further constrains on core overshoot were
obtained by the determination of the initial-final mass relationship, based on the
analysis of clusters turn-off and white dwarf cooling sequences \citep{cummings19}.

The inwards penetration of the bottom of the convective envelope during the RGB
evolution is a natural consequence of the expansion and cooling of the external
regions of the stars. This event, commonly known as first dredge-up (FDU), is easily
predicted by stellar evolution modelling, and leads to the modification of the
surface chemistry of the star, which can be tested spectroscopically. On the
other hand other processes might potentially alter the relative abundances of the
various chemicals, in a modality which is still highly debated \citep{hedrosa13, 
pinsonneault97, lagarde10}.

Capella ($\alpha$ Aurigae) is one of the most investigated and brightest binaries in 
the sky. Both components have estimated metallicity which is nearly solar \citep{fuhrmann11, torres15} and mass of the order of $2.5~{\rm M}_{\odot}$ \citep{weber11, torres15},
a mass range extremely interesting for a number of reasons: a) this mass is just above 
the threshold value separating the sources experiencing the helium flash from those 
undergoing quiescent helium burning; b) the study of the extension of the core mass 
during the MS phase of the star with mass close to those of Capella's
components proves very important, since the investigations developed so far have
shown that the trend of the extension of the extra-mixed region with mass exhibits 
a clear knee for masses below $\sim 2.5~{\rm M}_{\odot}$ \citep{claret17}; c)
the studies focused on dust production by low and intermediate mass stars 
outlined that the stars of mass in the $2.5-3~{\rm M}_{\odot}$ range are 
the most efficient dust manufacturers among the stellar sources, owing to
the large carbon enhancement achieved during the asymptotic giant branch
evolution, which triggers the production of large quantities of carbonaceous
dust \citep{ventura14, flavia15, flavia21}.

\begin{table}
\caption{Physical and chemical parameters of Capella}             % title of Table
\label{table}      % is used to refer this table in the text
\centering                          % used for centering table
\begin{tabular}{c c c c }        % centered columns (4 columns)
\hline\hline                 % inserts double horizontal lines
 & Primary & Secondary & Ref.  \\    % table heading 
\hline   
 Mass (${\rm M}_{\odot}$) & 2.5687$\pm$0.0074  & 2.4828$\pm$0.0067 & Torres \\
 T$_{\rm eff}$ (K) & 4970$\pm$50 & 5730$\pm$60 & Torres \\
 $\log ({g})$ (cgs)  & 2.691$\pm$0.041 & 2.941$\pm$0.032 & Torres \\
 \hline
  T$_{\rm eff}$ (K) & 4943$\pm$23 & 5694$\pm$73 & Takeda \\
  $\log ({g})$ (cgs) & 2.52$\pm$0.08 & 2.88$\pm$0.17 & Takeda \\
\hline
$^{12}$C$/^{13}$C & 17.8$\pm$1.9 & -  & Sablowski \\
C$/$N & 0.57$\pm$0.06 & 3.30$\pm$0.16 & Torres \\
$[$Na$/$H$]$ & -0.11$\pm$0.10 & +0.09$\pm$0.08 & Torres \\
$[$Na$/$H$]$ & +0.414 & -0.216 & Takeda \\ 
\hline
\end{tabular}
\tablefoot{Torres: \citet{torres15}, Takeda: \citet{takeda18}, Sablowski: \citet{sablowski19}.}
\end{table}

The study of Capella proves therefore particularly important because of the tight estimate of the 
mass of both components and of their physical properties, mainly surface gravity
and effective temperature \citep{torres15, takeda18}. Furthermore, the surface chemical
composition was determined through atmosphere models and
spectrum synthesis \citep{fuhrmann11, torres15, takeda18}.
%via detailed spectroscopic analysis \citep{weber11}, and by means of detailed model-atmosphere modelling \citep{takeda18}. 
The interpretation of
these information allows to deduce the extension of the convective core during the
MS evolution (which is crucial to reproduce the luminosity of both components) and of 
the extent of mixing from the base of the convective envelope during the FDU 
event (this is crucial to reproduce the observed chemical composition, particularly of the
primary, which has presumably already experienced the FDU). The latter point has received 
renewed interest in the last few years, owing to the recent study by \citet{sablowski19},
who used Large Binocular Telescope (LBT) and Vatican Advanced Technology Telescope (VATT), 
data to derive the surface carbon ratio of the primary star, obtaining a 
$^{12}$C$/^{13}$C$\sim 18-20$, significantly smaller than found in previous 
investigations \citep{tomkin76}. These observations were carried out with the Potsdam 
Echelle Polarimetric and Spectroscopic Instrument \citep[PEPSI;][]{strassmeier15} installed 
on the pier of LBT. The ultra high-resolution mode provided by PEPSI (R=250'000) was used 
to derive the isotope ratio with the use of spectrum synthesis from the CN lines at 8004 $\AA$.

In the present study we reproduce the surface gravities and effective temperatures of the
two components of Capella, and the surface chemical composition of the primary star (which are summarized in Table~\ref{table}), 
by comparing with results from detailed stellar evolutionary models, calculated ad hoc for the
two component of the system. The present analysis aims at the determination of the physical
ingredients, particularly the extent of the core overshoot during the MS phase, which allow
the best agreement with the observational evidence. The derivation of the age of the system 
will be the natural outcome of such an analysis. We will first consider the possibility that
the primary and secondary stars evolved at constant mass, equal to those determined nowadays;
we will also explore the case that the primary suffered mass loss while evolving during the 
RGB, and that formed with a higher mass than the current one.

The paper is structured as follows: the numerical and physical input used to calculate
the evolutionary sequences used for the present analysis are described in section
\ref{model}; a general description of the structural and evolutionary properties of the
stars with mass similar to those of Capella's components is given in section \ref{25msun};
in section \ref{capella} we reconstruct the evolutionary history of Capella, and discuss
the extent of the core overshoot allowing the best agreement with the observational evidence;
finally, the conclusions are given in section \ref{concl}.

\section{Stellar evolution modelling}
\label{model}
The evolutionary sequences used in the present investigations were calculated by means
of the ATON code for stellar evolution, in the version documented in \citet{ventura98},
where the interested reader can find the details of the numerical structure of the
code. The latest updates, including the nuclear network adopted and the micro-physics
used, are described in \citet{ventura09}. Here we briefly discuss the physical and chemical
ingredients of the code most relevant for the present study.

We considered model stars of initial mass in the $2.5-2.7~{\rm M}_{\odot}$ range, with solar
metallicity. The solar mixture was taken according to \citet{lodders}. The initial
metallicity and helium mass fraction were set to $Z=0.014$ and $Y=0.268$, respectively.

\subsection{The convective model}
\label{conv}
The location of the borders separating the regions of the stars characterized by convective motions
from the radiatively stable layers are found via the classic Schwarzschild criterion.
The temperature gradient with regions unstable to convective motions was calculated by means
the full spectrum of turbulence (FST) model \citep{cm91}. The mixing length within the deep
interiors of convective zones is assumed to be $\Lambda=z$, where $z$ is the distance from the nearest convective border; close the boundaries we use $\Lambda=z+\beta {\rm H}_p$; the free parameter was set to $\beta=0.2$, 
in agreement with the calibration of the solar model given in \citet{cm91}. Some further 
choices of $\beta$ were considered for the present work.

\subsection{Nuclear activity in convective regions}
\label{zeta}
Nuclear burning and mixing of chemicals are coupled with the diffusive schematization 
described in \citet{cloutman76}. The diffusive coefficient entering this
treatment is $D=1/3{\rm v}_c\Lambda$, where the convective velocity ${\rm v}_c$ is calculated by means of 
Eqs. 88, 89, 90 of \citet{cgm96}. For what attains mixing of chemicals beyond the formal borders, 
found via the Schwarzschild criterion, we consider an exponential decay of velocities, with a 
decay factor ${\rm exp}\big( {1\over {\zeta f_{\rm thick}}}\log {{\rm P}\over {\rm P_b}} \big)$. 
In the expression above ${\rm P_b}$ is the pressure at the formal boundary set by the
Schwarzschild criterion, $f_{\rm thick}$ is the thickness of the convective region in fractions 
of the pressure scale height ${\rm H}_p$, $\zeta$ is the parameter giving the e-folding distance of 
the convective velocity decay within radiatively stable regions, thus related to the extent of the 
extra-mixing occurring beyond the Schwarzschild border. The exponential approach to treat 
extra-mixing was applied also from the base of the convective envelope during the ascending of the
RGB.

We note that convective velocities
vanish at the formal border of convection. Therefore, as described in detail in section 2.2
of \citet{ventura98}, the exponential decay is started by a point within the convective
zone, where the pressure is $5\%$ different with respect to the pressure at the convective
boundary. As showed in \citet{ventura98}, assuming that the decay starts from points
where the pressure differences are $2\%$, $5\%$ and $10\%$ has no impact on the velocity profile.

The treatment adopted here is
physically equivalent to those used in other investigations \citep{herwig98, torres15} based on 
the exponential decay of velocities within radiative regions, as indicated by the numerical 
simulations by \citet{freytag96}. On the other hand, the
formalism is different, as in the present paper the decay is given as a function of pressure, 
whereas in the other papers the authors use the distance from the border: therefore the values
of $\zeta$ that will be derived here cannot be compared directly with the results of the other
investigations.

During the core He-burning phases we also considered diffusive extra-mixing, as
described above. We did not assume any specific treatment of semi-convection \citep{castellani71}
in the present work, as the overshooting assumption pushes the boundary of the mixing zone beyond 
the critical point where the instabilities associated to semi-convection might arise \citep{caloi93}.

\section{The evolutionary properties of Capella's components}
\label{25msun}

\begin{figure}
\begin{minipage}{0.48\textwidth}
\resizebox{1.\hsize}{!}{\includegraphics{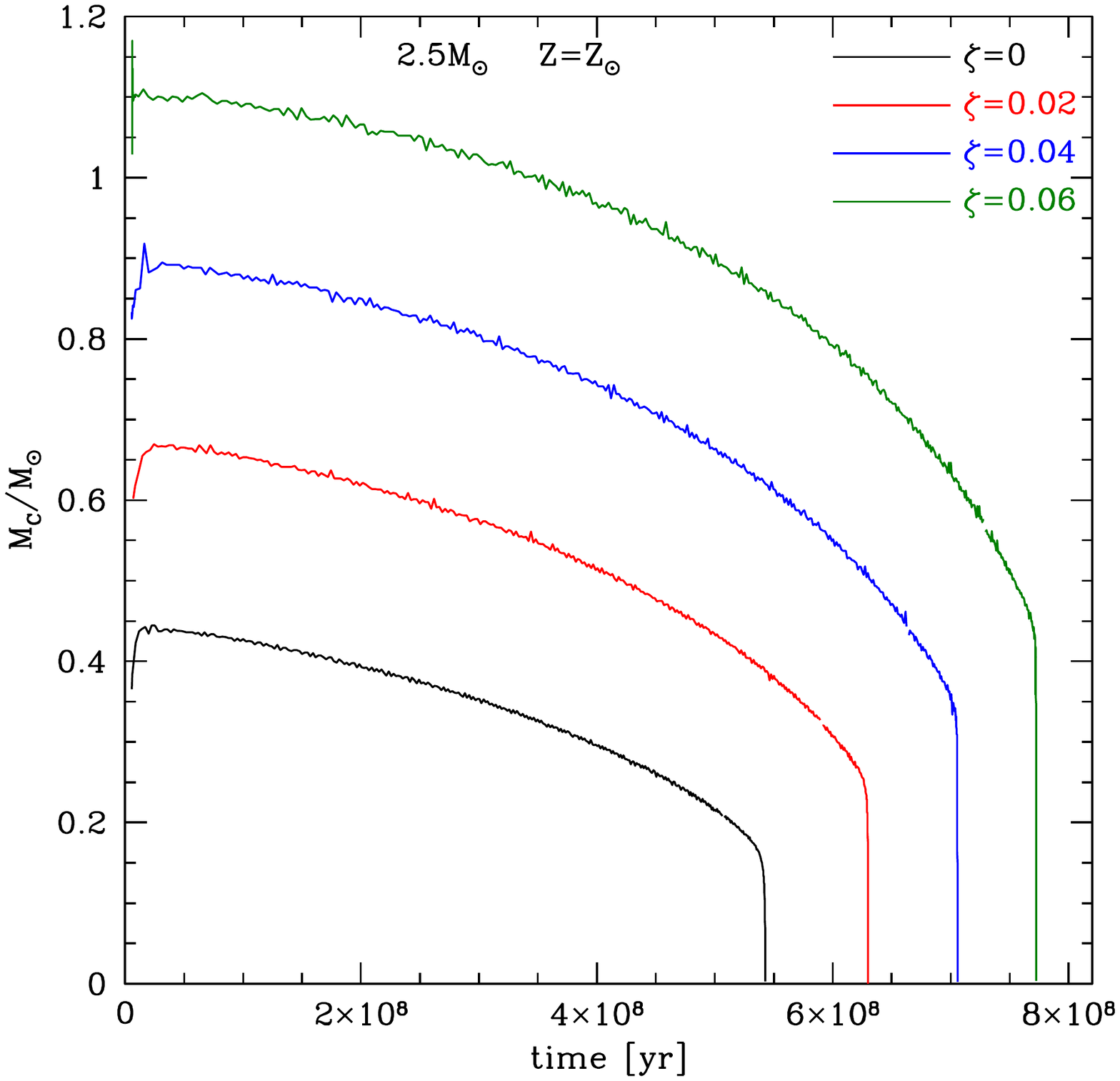}}
\end{minipage}
\vskip-50pt
\caption{Variation with the time of the mass of the central region of the star mixed by convective
currents, of a model star of initial mass $2.5~{\rm M}_{\odot}$ and solar metallicity during the hydrogen 
burning phase, adopting $\zeta$=0 (black line), $\zeta$=0.02 (red), $\zeta$=0.04 (blue) and $\zeta$=0.06 (green).
}
\label{fmcore}
\end{figure}

The primary and secondary components of Capella have masses around
$2.5~{\rm M}_{\odot}$. The current masses should not differ substantially from
the initial values; this holds in particular for the secondary component, which
is currently evolving through the initial part of the RGB phase, when no significant 
mass loss has occurred. Therefore, to understand the evolutionary properties of these 
stars, we focus on solar metallicity model stars of initial mass $2.5~{\rm M}_{\odot}$. 

These stars develop a convective core during the MS lifetime, thus the results
obtained are sensitive to the assumed extra-mixing from the external border of 
the core during the hydrogen burning phase, which in the present work is described
by the parameter $\zeta$, introduced in section \ref{zeta}. The choice of 
$\zeta$ affects the extension of the mixed region beyond the formal border
of the convective core during the MS, thus the duration of the H-burning phase and
consequently the evolutionary time scale of the model stars. 

This can be seen in Fig.~\ref{fmcore}, where the different lines indicate the
time variation of the mass of the region around the centre of the star where
the surface chemistry shows trace of nuclear processing, which can be
considered as the mass of the star within which convective mixing was sufficiently
efficient to carry the products of the nuclear activity occurring in the deep
interior. We find that the border of this region is located at the layer of the
star where the diffusion coefficient, defined in section \ref{zeta}, drops below 
$10^{-10}$ c.g.s. unit: the masses reported in Fig.~\ref{fmcore} refer to this layer. 
This choice implicitly implies that the definition of the 
masses shown in Fig.~\ref{fmcore} is not physical. However, the quantities 
shown in the figure can be safely considered to deduce the extent of the regions of the stars 
touched by convective currents. Furthermore, the comparison of the individual tracks with
the black one, which corresponds to the no-overshoot case, allows to infer the
extent of the overshoot zone, according to the choice of $\zeta$.

In all cases we recognize the
typical behaviour of the convective cores during the MS, which shrinks as hydrogen
is converted into helium, then vanish when the central hydrogen is exhausted.
The largest extension (in mass) of this region is $\sim 0.45~{\rm M}_{\odot}$ when 
no overshoot is assumed, and $\sim 1.1~{\rm M}_{\odot}$ when $\zeta=0.06$. 
Larger $\zeta$'s also imply longer time scales, as higher quantities of hydrogen are 
transported within the innermost, nuclearly active regions. The duration $\tau_{\rm H}$ 
of the H-burning phase changes from $\tau_{\rm H}=540$ Myr, for the no-overshoot case, 
to $\tau_{\rm H}=770$ Myr, for $\zeta=0.06$. Roughly, we find that $\delta \zeta=0.01$ 
corresponds to an age increase $\delta \tau_{\rm H} \sim 40$ Myr

\begin{figure*}
\begin{minipage}{0.48\textwidth}
\resizebox{1.\hsize}{!}{\includegraphics{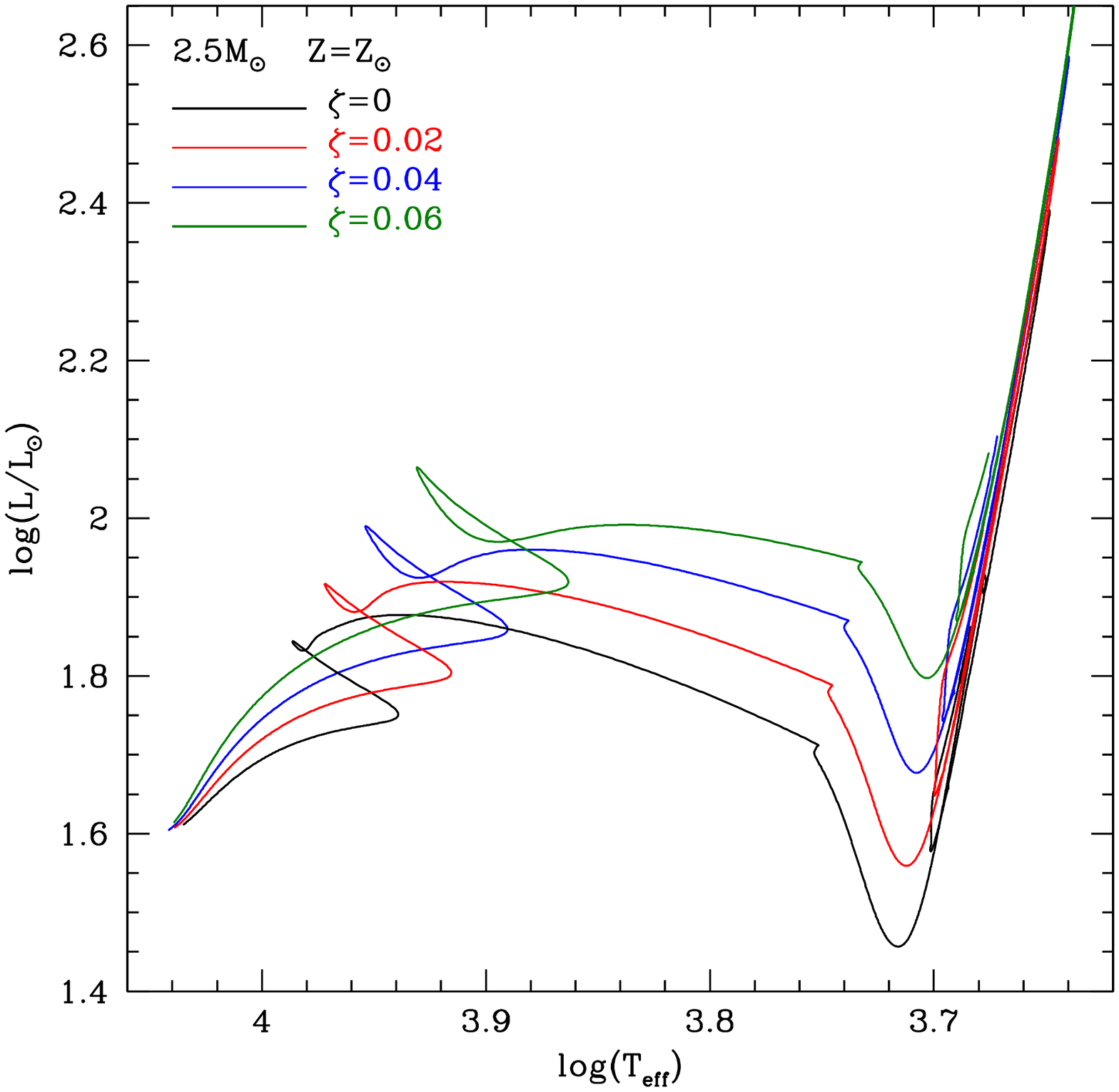}}
\end{minipage}
\begin{minipage}{0.48\textwidth}
\resizebox{1.\hsize}{!}{\includegraphics{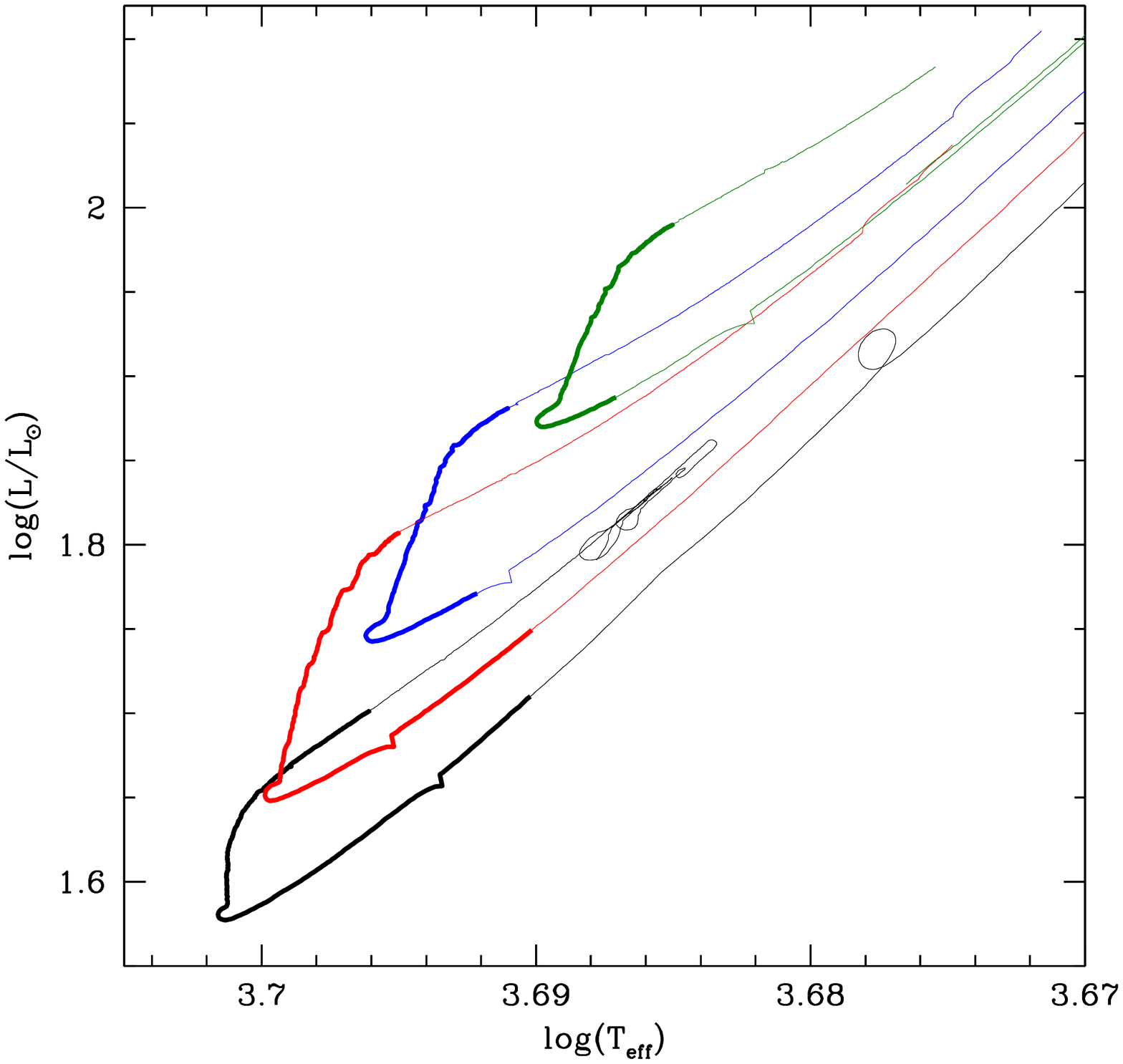}}
\end{minipage}
\vskip-50pt
\caption{Left: Evolutionary track of 2.5${\rm M}_{\odot}$ model star with solar metallicity on the HR diagram, calculated with the same values of $\zeta$ as Fig.~\ref{fmcore}. Right: Same as left panel, but focusing on the clump region, where the slowest part of the evolution is highlighted by a thick line.}
\label{fhr}
\end{figure*}

The left panel of Fig.~\ref{fhr} shows the evolutionary tracks on the HR diagram of 
the $2.5~{\rm M}_{\odot}$ model stars reported in Fig.~\ref{fmcore}. The choice of 
$\zeta$ affects the morphology of the tracks in the region of the turn-off, for the reasons 
discussed earlier in this section, and the luminosity of the clump described by the tracks 
during the core helium burning phase. This is more clear in the right panel of 
Fig.~\ref{fhr}, which is zoomed on the clump region. The thickest part 
of each track corresponds to the slowest part of the evolution, accounting for $90\%$ of 
the overall post H-burning time.
The luminosity range spanned by the clump is $40-50~{\rm L}_{\odot}$ for the no-overshoot case,
$45-65~{\rm L}_{\odot}$ for $\zeta=0.02$, $55-75~{\rm L}_{\odot}$ for $\zeta=0.04$, and
$75-95~{\rm L}_{\odot}$ for $\zeta=0.06$.

\begin{figure}
\begin{minipage}{0.48\textwidth}
\resizebox{1.\hsize}{!}{\includegraphics{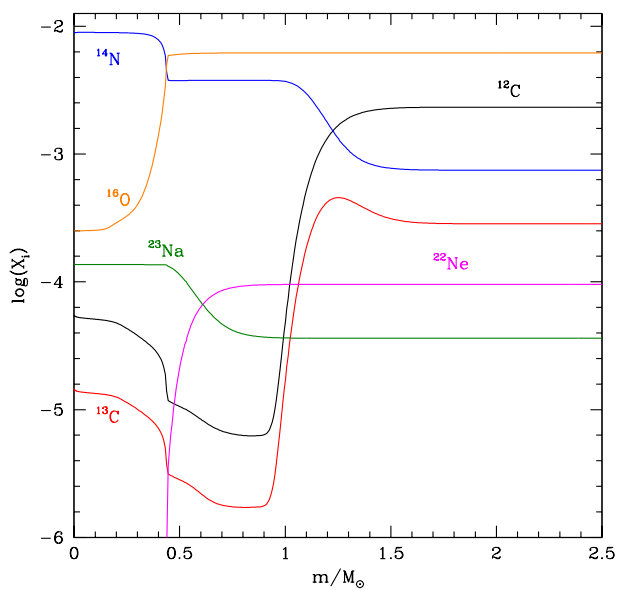}}
\end{minipage}
\vskip-50pt
\caption{Internal chemical stratification of a model star of 2.5${\rm M}_{\odot}$ with solar metallicity at the end of the core hydrogen burning phase, calculated by assuming no overshoot ($\zeta$=0).
}
\label{fchem}
\end{figure}

The nuclear activity during the MS lifetime alters the chemical
composition of the internal regions of the star, which is modified according 
to the proton capture nucleosynthesis experienced. The chemical stratification
at the end of the MS turns extremely important to understand the change
in the surface chemistry of the star taking place as a consequence of the
FDU. Fig.~\ref{fchem} shows the mass fractions of various
chemical species at the end of the core H-burning phase. The figure
refers to the $2.5~{\rm M}_{\odot}$ model star calculated with $\zeta=0$
(black lines in Figg. \ref{fmcore} and \ref{fhr}).

The whole ${\rm M}<1.5~{\rm M}_{\odot}$ region of the star is touched by
the internal nucleosynthesis. We recognize the typical double step in the
nitrogen profile, indicating the activation of the full CNO cycle for
${\rm M}<0.45~{\rm M}_{\odot}$, and the plain CN cycle in the more external
part of the core. This is confirmed by the depletion of the oxygen, which in the
center of the star is depleted by a factor $\sim 20$ with respect to the initial
quantity. Regarding the carbon isotopes, $^{12}$C is depleted within
the whole internal region of the star, whereas a peak in the $^{13}$C profile
is present at ${\rm M}\sim 1.2~{\rm M}_{\odot}$; both $^{12}$C and $^{13}$C are heavily
depleted by more than a factor 100 in the innermost layers. 

The chemistry of the region ${\rm M}<1~{\rm M}_{\odot}$ is also affected by 
Ne-Na cycling, with the depletion of $^{22}$Ne and the synthesis of sodium, 
whose central abundance exceeds the surface one by a factor $\sim 4$.

\begin{figure}
\begin{minipage}{0.48\textwidth}
\resizebox{1.\hsize}{!}{\includegraphics{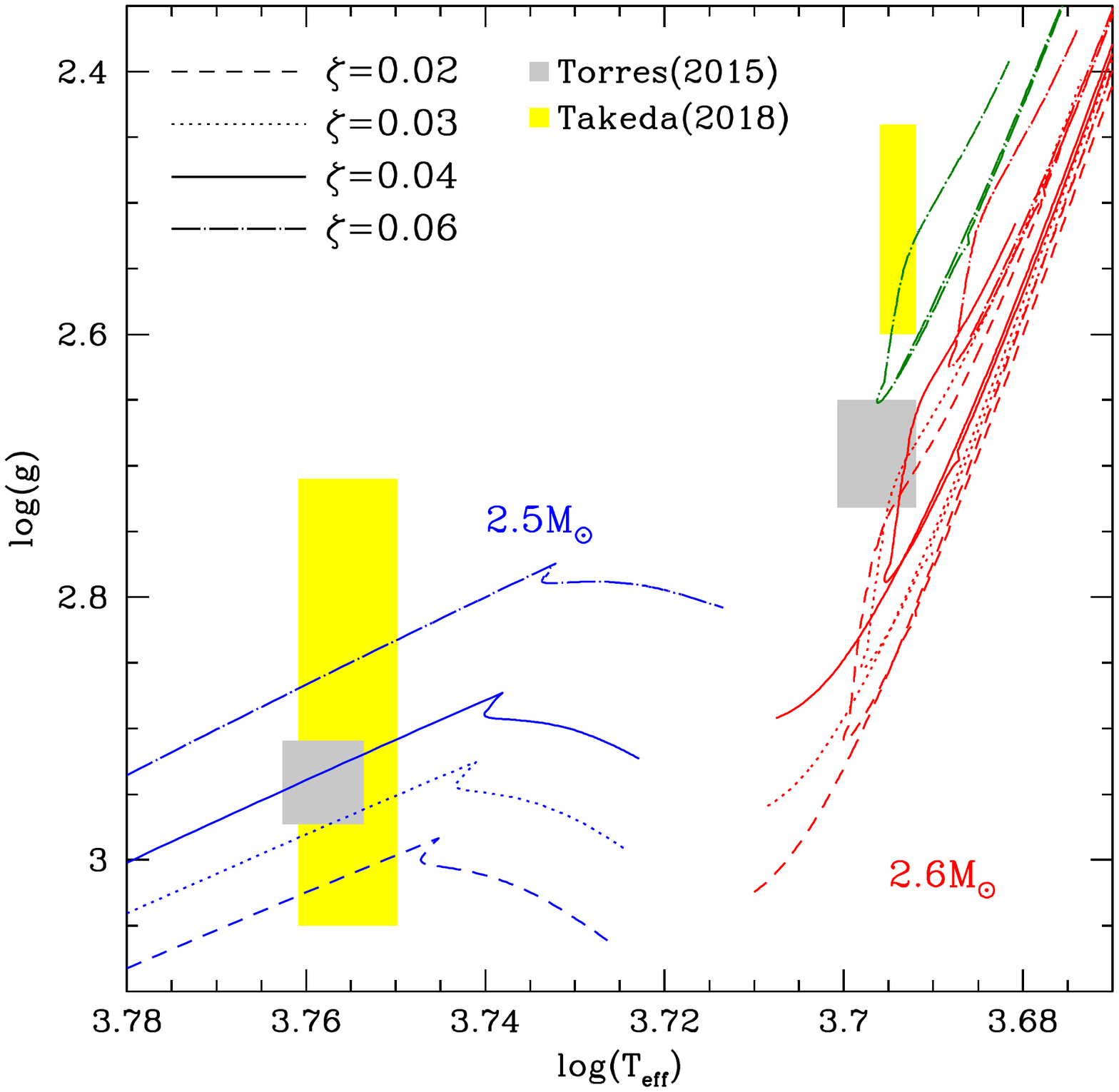}}
\end{minipage}
\vskip-50pt
\caption{Evolutionary tracks on the $\log({\rm T}_{\rm eff})-\log({\rm g})$ plane of model stars of mass 
$2.5~{\rm M}_{\odot}$ (blue lines) and $2.6~{\rm M}_{\odot}$ (red), calculated by assuming 
different values of the overshoot parameter $\zeta$. The green line refers to
a $2.6~{\rm M}_{\odot}$ model star obtained by assuming $\zeta=0.06$ and an FST parameter
$\beta=0.27$. Grey and yellow rectangles indicate the
error boxes for the two components of Capella given, respectively, in \citet{torres15} and
\citet{takeda18}.
}
\label{figtg}
\end{figure}

\section{Understanding the evolution of Capella}
\label{capella}
To reconstruct the past history of Capella we use evolutionary tracks calculated 
on purpose for the present investigation, for the primary and the secondary components, using the 
input described in section \ref{model}. The goal of this analysis is the determination of the 
combination of the parameters $\zeta$ and $\beta$, described in section \ref{zeta} and
\ref{conv}, which allow to reach consistency between the results from modelling and those
from the observations. 

As a first step we require that the evolutionary tracks of the model stars
in the effective temperature - gravity plane reproduce the results obtained from the observations.
We will consider first the study by \citet{torres15}, then the more recent work by \citet{takeda18}:
the corresponding observational boxes for the two components of the system are shown in grey and yellow, respectively, in Fig.~\ref{figtg}. The fit of the surface gravities will lead to select the
appropriate range of values of $\zeta$. $\beta$ affects mainly the effective temperature, thus possible
changes with respect to the value required to reproduce the evolution of the Sun will be considered
only in the case that the effective temperatures of the stars are not reproduced by modelling.

The second step of this analysis will be to further restrict the range of parameters to be
considered, by requiring that the model stars of the primary and secondary evolve through the
observational boxes shown in Fig.~\ref{figtg} at the same time. This will also allow
the determination of the age of the system.

The final step of this procedure will be to check consistency between the
expected change of the surface chemistry, driven by the FDU, and the measured surface
abundances of the various chemical species.

\subsection{The determination of core extra-mixing}

Based on the high highly precise ($\sim$ 0.3) masses obtained by \citet{torres15} 
and reported in Table~\ref{table}, we first start by considering model stars of mass 
$2.6~{\rm M}_{\odot}$ and $2.5~{\rm M}_{\odot}$ for the primary and secondary star, 
respectively; this choice implicitly assumes that negligible mass loss occurred 
during the previous evolutionary phases. We will also consider the possibility that 
the primary star was initially more massive than nowadays, thus invoking some mass 
loss during the RGB ascending.

Fig.~\ref{figtg} shows the evolutionary tracks of model stars of mass $2.5~{\rm M}_{\odot}$ 
and $2.6~{\rm M}_{\odot}$, calculated with different $\zeta$'s. The observational
boxes for both components of Capella, as found by \citet{torres15}, are shown as
grey rectangles. From the results shown in 
Fig.~\ref{figtg} we deduce that only $\zeta$'s in the 0.03-0.04 range can
be considered: values of $\zeta$ external to this range do not allow to fit the
surface gravity (hence the luminosity) of both stars.

\begin{figure*}
\begin{minipage}{0.48\textwidth}
\resizebox{1.\hsize}{!}{\includegraphics{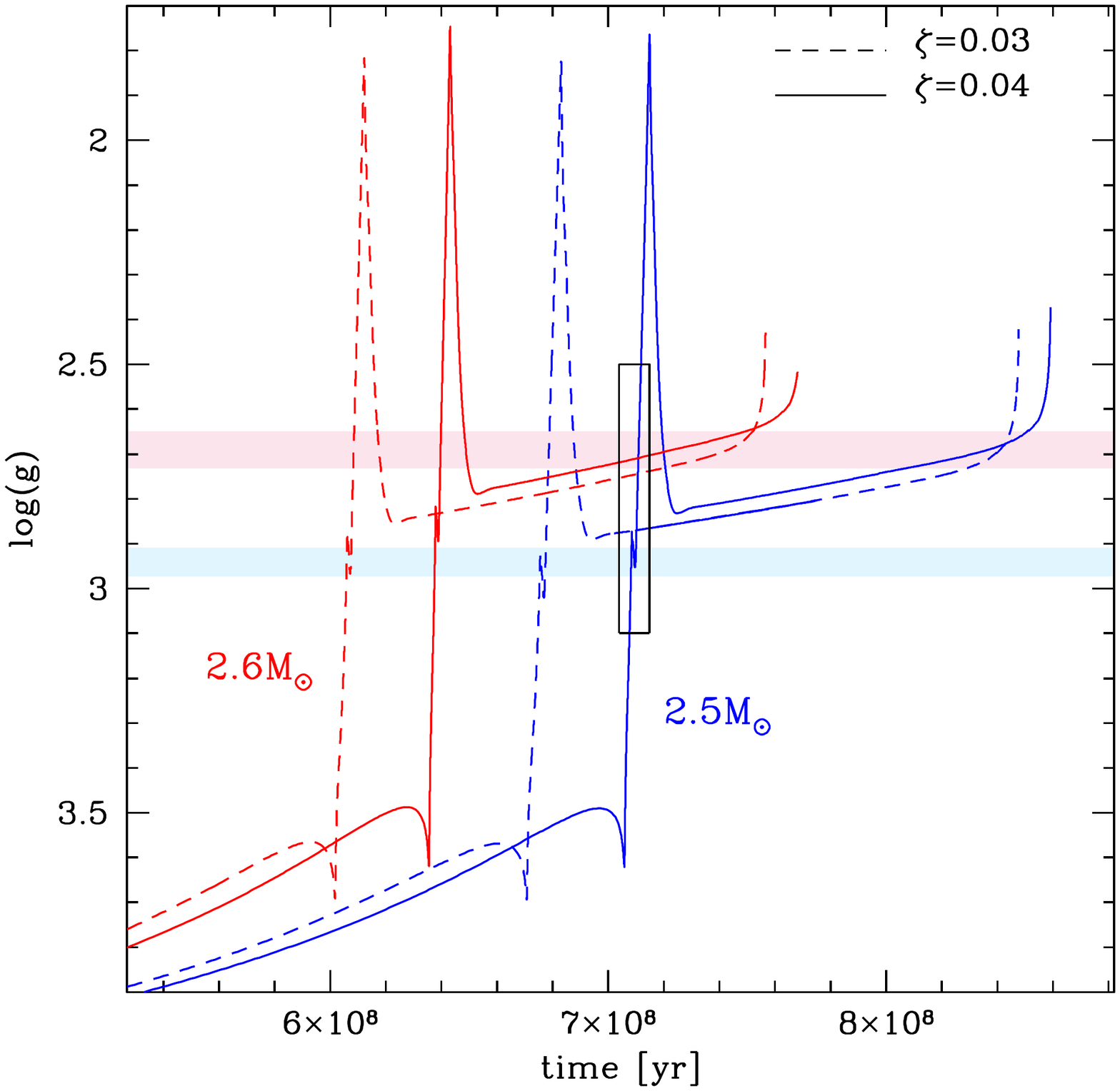}}
\end{minipage}
\begin{minipage}{0.48\textwidth}
\resizebox{1.\hsize}{!}{\includegraphics{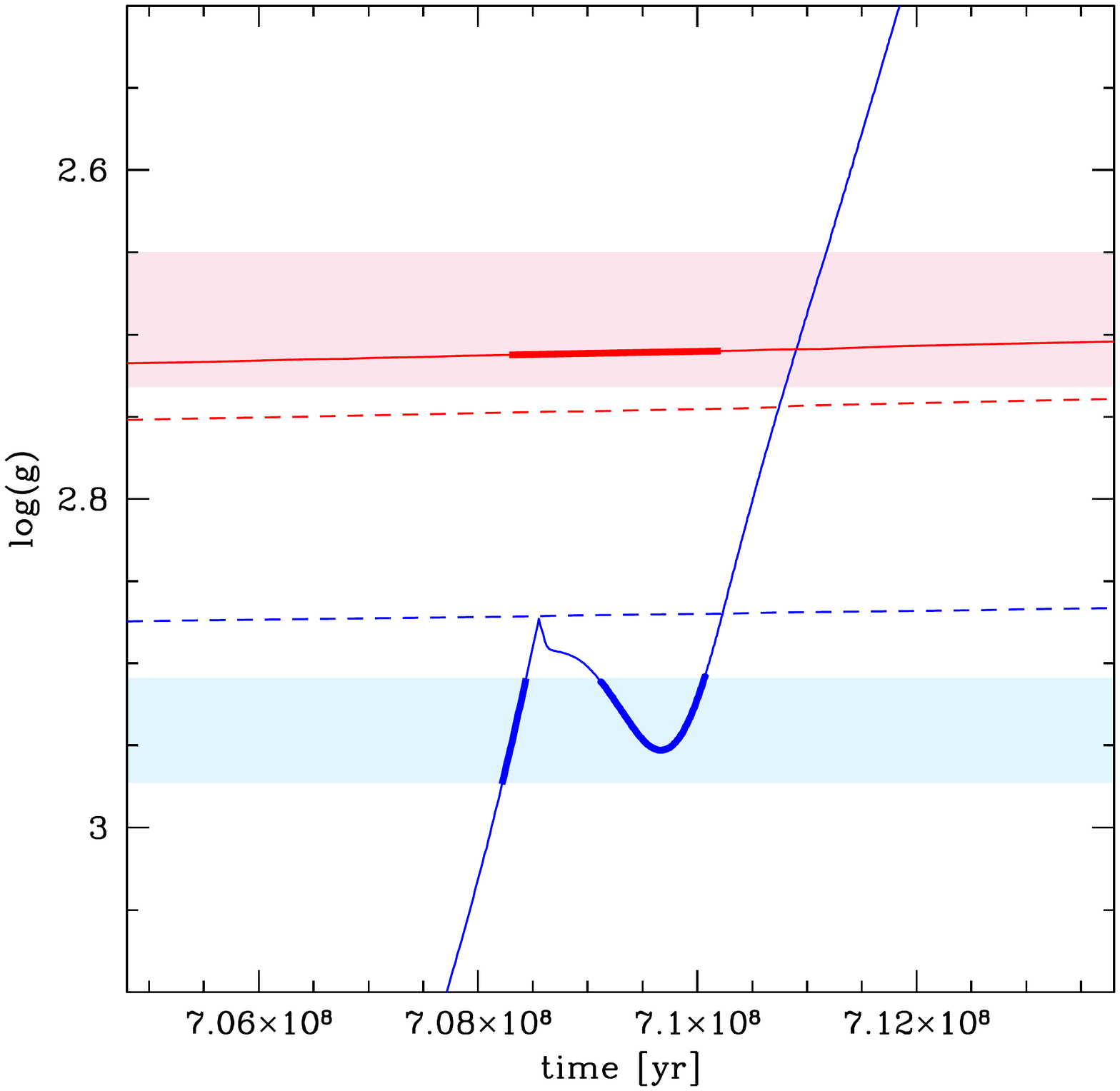}}
\end{minipage}
\vskip-50pt
\caption{Time variation of the $\log(g)$ of model stars of mass $2.5~{\rm M}_{\odot}$ (blue lines) and $2.6~{\rm M}_{\odot}$ (red), calculated by assuming $\zeta=0.03$ and $\zeta=0.04$. Coloured regions indicate the error associated to the derivation of the surface gravity for the two components of Capella (pink - primary, cyan - secondary) given in \citet{torres15}. The black box on the left panel is zoomed on the right, in which the thick lines highlight the simultaneous fit of both components.}
\label{fagegrav}
\end{figure*}

As discussed previously, we also require that the two stars evolve through the 
respective observational box at the same epoch. In Fig.~\ref{fagegrav} we show the 
evolution of the surface gravity of different model stars, calculated 
with different values of $\zeta$. The surface gravities derived by 
\citet{torres15} are shown as coloured regions. 
We see in Fig.~\ref{fagegrav} that consistency is found in the $\zeta=0.04$ case, which leads 
to an estimated age of 710 Myr. This is more evident in the right panel of 
Fig.~\ref{fagegrav}, which is focused on the region inside the black box. 
We note that this choice corresponds to an assumed instantaneous extra-mixing from the 
core of $\sim 0.25~{\rm H}_p$. 

On the other hand, adopting $\zeta=0.03$ does not allow the 
simultaneous fit of both components: indeed in this case the age derived on the basis of 
the modelling of the secondary star, i.e. 670 Myr, is at odds with the age derived 
for the primary, as the latter star is expected to reach the observed gravities 
only during the final part of the clump evolution, at an age of 710 Myr.

\begin{figure}
\begin{minipage}{0.48\textwidth}
\resizebox{1.\hsize}{!}{\includegraphics{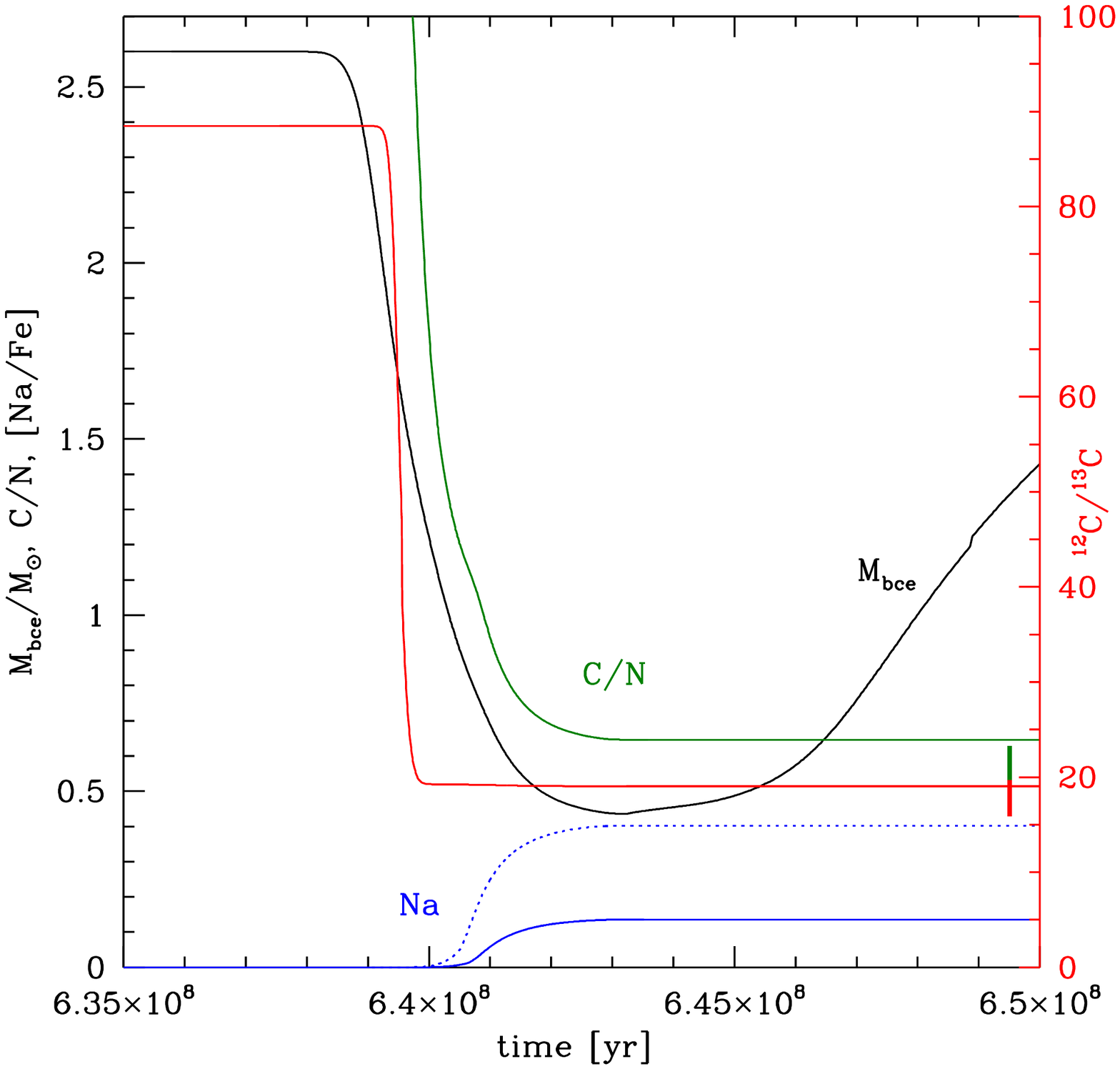}}
\end{minipage}
\vskip-50pt
\caption{Time variation of the mass at the bottom of the convective
envelope (black, scale on the left), of the surface sodium (blue,
scale on the left), and of the carbon isotopic ratio (red, scale 
on the right) of a $2.6~{\rm M}_{\odot}$ model star, calculated
with an overshoot parameter $\zeta=0.04$. The plot is focused on
the epoch during which the FDU takes place. The dashed,
blue line indicates the variation of the surface sodium abundances
of a model star calculated by assuming that the initial neon is
enhanced by a factor 3. 
}
\label{f1dup}
\end{figure}

For what concerns the choice of the parameter $\beta$ entering the expression
of the mixing length in the FST modelling, the results reported in Fig.~\ref{figtg}
suggest that adopting $\beta=0.2$ allows to reproduce the effective temperatures
derived for both the components of Capella, although a slightly higher value 
cannot be ruled out on the basis of the present analysis. This is the first
confirmation, to date, that the same quantity originally derived by \citet{cm91}
to reproduce the evolution of the Sun can be applied in a different physical
situation, such as the convective envelope of a star evolving through the RGB
and the core helium burning phase.

According to the present analysis Capella is $\sim 80$ Myr younger than
found by \citet{torres15}, based on the PARSEC isochrones of \citet{bressan12}. 
The reason for this difference is mainly due to the different chemistry adopted:
\citet{torres15} use a slightly sub-solar metallicity and initial helium 
${\rm Y}=0.272$, whereas we use solar metallicity and helium ${\rm Y}=0.268$.
Both the higher metallicity and the lower helium adopted here favour older ages,
of the order of $\sim 40$ Myr. Further differences with respect to the PARSEC
models, which explain the residual age difference between us and \citet{torres15},
are the solar mixture adopted (we use the mixture by \citet{lodders}, whereas
\citet{bressan12} assume the element distribution by \citet{gs98}), and the 
prescription for core overshoot.

\subsection{A test of mass-loss during the RGB}
We also considered the possibility that the primary star experienced mass loss during 
the previous evolutionary phases, so that it descends from a higher mass progenitor. 
On the numerical side, when a plain mass loss description such as Reimers' is 
adopted, the decrease in the total mass by the time that the star reaches the core 
He-burning phase is within $0.01~{\rm M}_{\odot}$, for any reasonable choice of the 
free parameter entering the Reimers' recipe. However, we decided to investigate the 
effects of some mass loss, even significantly higher than found via the Reimers' 
description.

we first tested the case that the initial mass
was $2.7~{\rm M}_{\odot}$. In this case we find consistency with the temperature
and gravity derived from the observations of the primary, by assuming $\zeta=0.035$. 
However, this choice would pose severe issues on the side of the evolutionary times, 
as the secondary star would attain the corresponding observational quantities 
$\sim 50$ Myr ahead of the primary. To maintain simultaneity we should claim that the 
overshoot from the core of the secondary star was smaller, of the order of $\zeta=0.02$; 
however, as discussed earlier in this section (see also Fig.~\ref{figtg}), this choice is 
ruled out, as it does not allow to fit the surface gravity of the secondary component, 
given in \citet{torres15}.

\subsection{The study by \citet{takeda18}}
In a recent paper \citet{takeda18} published a new study on Capella, where the elemental
abundances were estimated by the disentangled spectrum of the two components.
The effective temperatures found by \citet{takeda18} are in substantial agreement
with those given in \citet{torres15}, with the only difference that the error associated
to the determination of the temperature of the primary, 23 K, is significantly lower than
the errors reported in the study by \citet{torres15}. The comparison between the
surface gravities derived in the two studies outlines that the gravities derived 
in \citet{takeda18} are significantly lower than those given by \citet{torres15}.
While in the case of the secondary star consistency is found when the error bars
are taken into account, in the case of the primary there is a difference 
$\delta (\log g) = 0.17$ between the two studies, which can be reduced at most
to $\delta (\log g) = 0.05$ when the corresponding error bars are considered.

To reproduce the surface gravities given by \citet{takeda18} we calculated new
evolutionary sequences for both the components of Capella, with 
values of $\zeta$ higher than those previously discussed: this is because
the lower gravities of the primary imply higher luminosities. We note that 
given the large error associated to the 
derivation of the surface gravity of the secondary star, this part of the 
study was essentially focused on the fit of the observed parameters of the
primary. 

Consistency with the surface gravity of the primary star 
is found when a $2.6~{\rm M}_{\odot}$ is evolved with $\zeta=0.06$, which
corresponds to an instantaneous overshoot from the outer border of the
convective core during the H-burning phase of $\rm {0.36~H_{P}}$. As shown 
in Fig.~\ref{figtg}, in this case the effective temperature during the 
helium burning phase, is slightly cooler with respect to the values
given in \citet{takeda18}, which required to use a higher FST parameter:
the choice $\beta=0.27$ allows a nice fit of the results from the observations
(green track in Fig.~\ref{figtg}).

\subsection{Which will be the final mass of the primary component of Capella?}
The main difference between the conclusions drawn when using the gravities
given in \citet{torres15} and those by \citet{takeda18} is the extension of the
extra-mixed zone during the MS evolution: in the former case we find $\zeta=0.04$,
which corresponds to a $0.25~{\rm H_P}$ size, whereas in the latter case we derive
$\zeta=0.06$, which corresponds to $0.37~{\rm H_P}$. 

To test the possibility of discriminating on the basis of the initial-final mass
relationship derived by \citet{cummings19}, we extended the computations of model stars of 
mass $2.6~{\rm M}_{\odot}$ until the first thermal pulse, thus at the beginning of the 
asymptotic giant branch (AGB) phase. The purpose of this test was to check whether either of
the two values given above leads to inconsistency with the \citet{cummings19} law.

In the $\zeta=0.04$ case we find a core mass of $0.53~{\rm M}_{\odot}$, whereas
when $\zeta=0.06$ is adopted we find $0.57~{\rm M}_{\odot}$. Unfortunately this
difference prevents any possibility of discriminating between the two scenarios.
As shown in \citet{ventura98}, model stars that attain core masses at the beginning 
of the AGB evolution similar to those mentioned above end up the AGB phase with a final 
mass of $\sim 0.7~{\rm M}_{\odot}$, which is anyway consistent with the expectations
based on the study by \citet{cummings19}.

\subsection{The surface chemical composition}
In Fig.~\ref{f1dup} we describe the evolution of the primary star
during the evolutionary phases close to the occurrence of the
FDU. The figure refers to the $2.6~{\rm M}_{\odot}$
model star calculated with $\zeta=0.04$, discussed earlier in this
section. The quantities shown are the mass of the
region of the star below the bottom of the convective envelope,
the surface sodium and the $^{12}$C$/^{13}$C and C$/$N ratios.

During the FDU the convective envelope penetrates inwards and 
grows in mass, up to including $\sim 2~{\rm M}_{\odot}$ of the star. 
The base of the convective mantle reaches regions of the star
whose chemical composition was altered by nuclear activity; the 
typical stratification of the star at the end of the core
hydrogen burning phase, before the occurrence of the FDU, is
shown in Fig.~\ref{fchem}. The occurrence of FDU favours the increase 
in the surface nitrogen and sodium, the decrease in the surface carbon, 
with an overall increase in $^{12}$C$/^{13}$C.

The latter quantity proved a severe issue in the attempt of reconstructing
the previous history of Capella, because the value given by \citet{tomkin76},
$^{12}$C$/^{13}$C$ \sim 27$, was significantly higher than that found from
stellar evolution modelling. The agreement between the theoretical expectations
and the observational results is now possible thanks to the recent work by 
\citet{sablowski19}. The authors, based on LBT and VATT observations carried out 
with PEPSI, found that the surface $^{12}$C$/^{13}$C of the primary star is 
$17.8 \pm 1.9$, significantly lower than that found by \citet{tomkin76}.
The reason for this difference is likely the fact that \citet{tomkin76} used
an incorrect light ratio between the two components, thus underestimating the 
equivalent width measurements of the CN lines by $6-11 \%$ \citep{torres09}; 
on the other hand, the analysis by \citet{sablowski19} benefited from the 
remarkable performances of PEPSI, which made it possible to obtain ultra-high 
resolution spectra with high signal-to-noise ratio ($\sim$2000), thus allowing an 
estimation of $^{12}$C$/^{13}$C of the primary star of Capella with unprecedented reliability. 
As shown in Fig.~\ref{f1dup}, the new measurement is in nice agreement with results 
from stellar evolution modelling.
%The improvement of the data, the synthesis analysis, and better spectral line data, allowed a more reliable estimation of $^{12}$C$/^{13}$C of the primary star of Capella,  
This result indicates that the surface chemistry of the primary star was
altered by the effects of the FDU only, with no additional contribution from some
other mechanisms, often invoked to occur during the RGB phase, to explain, e.g.,
the low $^{12}$C$/^{13}$C detected in open cluster stars \citep{gilroy91}. This is
consistent with the predictions by \citet{cl10}, who find that neither thermohaline
nor rotational mixing alter substantially the surface chemistry of stars in the mass 
domain explored here, during the RGB phase.

Concerning the C$/$N ratio, as shown in Fig.~\ref{f1dup}, we find that the expected 
value ($\sim 0.65$) is slightly in excess of the value reported in \citet{torres15}, C$/$N $\sim$ 0.57:
this might be a consequence of the fact that the lines used to estimate the C and N 
abundances do not fully represent the abundances in the photosphere \citep[see 
discussion in][on this argument]{torres15}.

Regarding the surface sodium of the primary, the results from \citet{torres15}
and \citet{takeda18} differ significantly: while according to \citet{torres15} the
star shows no sign of Na enrichment, \citet{takeda18} find that the primary
is sodium enriched by 0.4 dex. The latter result can be hardly explained within the
context of the FDU, as the expected increase in the surface sodium is
$\delta[$Na$/$Fe$]\sim 0.2$ (see Fig.~\ref{f1dup}). The assumption of extra-mixing
is of little help here, as the consequent increase in the final $[$Na$/$Fe$]$ is
within 0.05 dex. Among the various possibilities, we take into account that the
neon is underestimated, such that the initial neon is higher than that proposed
by \citet{lodders}. We explored various possibilities, and found that a sodium
enrichment in agreement with the results from \citet{takeda18} is obtained by assuming
an initial neon a factor 3 higher that in \citet{lodders}. A discussion regarding the
initial neon of solar metallicity stars is definitively beyond the scopes of the present 
investigation; however, we point out that the neon enhancement invoked here, of the order of
+0.5 dex, is consistent with the conclusions reached by \citet{bahcall05}, where a higher
neon is required to reconcile the predictions from solar modelling with heliosismological
measurements. We leave this problem open.

The results shown in Fig.~\ref{f1dup} were obtained by assuming extra-mixing
from the convective core during the H-burning and from the bottom of the envelope
during the RGB phase. The latter assumption has no effect on the evolutionary time
scale and on the excursion of the evolutionary track across the HR diagram, but might
potentially alter the extent of the inwards penetration of the envelope, thus the
change in the surface chemical composition following the occurrence of the FDU.
To this aim we calculated further evolutionary sequences, where we explored different
values of the $\zeta$ parameter in the $0-0.06$ range to describe the overshoot from 
the base of the surface convection; for the core we used $\zeta=0.04$ in all cases.

In the no-overshoot case we find differences in the final abundances of $^{12}$C, $^{13}$C
and $^{14}$N of the order of $10\%$, while the final sodium is $15\%$ lower than for the
$\zeta=0.04$ case. In this case the $^{12}$C$/^{13}$C and C$/$N ratio are significantly higher
than those resulting from the spectroscopic analysis, whereas the discrepancy in the
surface sodium between theory and the observations is amplified.
On the other hand, when $\zeta$ is chosen in the $0.02-0.06$ range,
the results turn out almost independent of $\zeta$. The reason for this is that for any choice
of $\zeta \geq 0.02$ the base of the mixed regions approaches the H-burning shell, thus 
convection is pushed outwards.

This exploration shows that some extra-mixing from the base of the convective zone is
required to reproduce the observations, although it is not possible to assess the extension of
the extra-mixed zone in detail.

%-----------------------------------------------------------------

\section{Conclusions}
\label{concl}
The evolutionary history of stars forming the binary system $\alpha$ Aurigae, or
Capella, is reconstructed, by means of the comparison between results from stellar evolution 
modelling and those derived from observations. The conclusions reached in this
work have been mostly based on the analysis of the primary star, which is likely 
evolving through the core helium burning phase, and has already experienced the 
change in the surface chemical composition favoured by the occurrence of the first
dredge-up; conversely, the secondary star is currently evolving through the early phase
of the red giant branch, thus the derived luminosity and chemical composition turn out of
little help in this context.

The results from the present analysis indicate that Capella formed $\simeq$710 Myr ago. 
The comparison between the
derived surface gravity and effective temperature of the primary star confirm that
extra-mixing from the convective core occurred during the main sequence phase, with 
the extension of the extra-mixed zone being of the order of $0.25~{\rm H_P}$. 
This result is in agreement with previous works focused on the determination of the size of 
the convective cores of stars of mass similar to those of Capella's components.
Regarding the full spectrum of turbulence convection modelling used here, we find that
the expression for the mixing length required to reproduce the evolution of the Sun
is consistent with the effective temperatures found for the primary and secondary stars
of Capella.

The estimate of the core overshoot must be reconsidered upwards if we consider a more recent, lower estimate of 
the surface gravity (hence a higher luminosity) of the primary component, which leads to a 
larger size of the overshoot zone, of the order of $0.37~{\rm H_P}$.

The nowadays observed surface chemistry of the primary star can be explained
as a result of the occurrence of the first dredge-up only: we confirm previous
results from detailed stellar evolution modelling, that stars of mass 
$\sim 2.5~{\rm M}_{\odot}$ during the RGB phase experience no additional mixing 
mechanisms, such as thermohaline and rotational mixing, likely taking place in 
the lower mass counterparts. This conclusions relies on the observed 
C$/$N and $^{12}$C$/^{13}$C ratios, the latter having been recently found using high-resolution spectra obtained with PEPSI at LBT and VATT.  Recent results from detailed model-atmosphere analysis of the disentangled spectrum outline a significant sodium enhancement, of the order of 0.4 dex, which we tentatively explain by invoking a higher neon content of the star.

\begin{acknowledgements}
      EM is indebted to Guillermo Torres and Yoichi Takeda for the helpful discussions.
      EM acknowledges support from the INAF research project “LBT - Supporto Arizona Italia". MT acknowledges support from the ERC Consolidator Grant funding scheme (project ASTEROCHRONOMETRY, https://www.asterochronometry.eu, G.A. No. 772293).
\end{acknowledgements}


\end{document}